# PROPER MOTION OF REFERENCE RADIO SOURCES

O. Titov
Geoscience Australia
PO Box 378 Canberra 2601 Australia
e-mail: oleg.titov@ga.gov.au

The motion of relativistic jets from the active extragalactic nuclei can reach several hundred microseconds per year and mimic proper motion of the distant radio sources observed by VLBI. Such motion of individual quasars is not correlated and its magnitude exceeds the small systematic effects induced by the rotation of the Solar system around the centre of the Galaxy. In this paper we search for the cause of the systematic effect and discuss the results.

## INTRODUCTION

Multi-frequency VLBI measures accurate positions of reference extragalactic radio sources. Several hundred such sources (mostly quasars and radio-galaxies) have their positions known to an accuracy of better than 0.5 mas with respect to the Solar system barycenter and used to construct the International Celestial Reference Frame (Ma et al, 1998). Apparent proper motion of these sources, presumably induced by the source structure changes, reaches several hundred µas/year for selected radio sources (Ma et al. 1998, Fey et al. 2004). These variations of the radio source structure are supposed to have random orientation therefore the apparent proper motion should not be systematic. Due to their large distances, transverse proper motion caused by the Solar system motion is negligible (Sovers et al, 1998). However, systematic effects in the radio source positions can appear due to a variety of reasons, for instance, secular aberration drift, gravitational waves and the Hubble constant anisotropy, and in principle, could be used to provide information about various cosmological processes.

## SECULAR ABERRATION DRIFT, GRAVITATIONAL WAVES AND THE HUBBLE CONSTANT ANISOTROPY

On the Galactic scale, the Solar system barycenter rotates around the center of mass of the Galaxy with period about 200 million years (Saslaw, 1985). Centrifugal galactocentric acceleration of the Solar system barycentre results in the appearance of another effect of special relativity (rate of change of secular aberration with time – "secular aberration drift") that systematically changes estimates of apparent positions of all celestial bodies (Gwinn et al., 1997, Sovers et al., 1998; Kovalevsky, 2003; Klioner, 2003; Kopeikin and Makarov, 2005). For the distance of the centre of the Galaxy of 10 kpc the magnitude of the galactocentric acceleration vector is about 5 µas/year and directed towards the center of the Galaxy. If the origin of the celestial reference frame based on quasars is co-incident with the Solar system barycenter, positions of these quasars with respect to the Solar system barycenter should change in a form of additional systematic proper motion.

In a case of shear-free isotropic expansion of the Universe,

$$V = HR \tag{1}$$

($V$ – radial velocity, $R$ – distance to the quasar, $H$ – the Hubble constant) the Hubble constant is uniform around the sky, so any transverse proper motion vanishes. However, if the Universe expansion is anisotropic (though still shear-free), then the Hubble law for the isotropic Universe (1) should be replaced by

$$V = \left[ e_{33} \sin^2 \delta + \frac{1}{2}(e_{11} + e_{22}) \cos^2 \delta + \frac{1}{2}(e_{11} - e_{22}) \cos 2\alpha \cos^2 \delta \right] R = \\ = (H + \Delta H_3 \sin^2 \delta + \Delta H_{12} \cos 2\alpha \cos^2 \delta) R \quad (2)$$

where $e_{11}, e_{22}, e_{33}$ - diagonal elements of the expansion tensor, ($\alpha$, $\delta$) – equatorial coordinates. The Hubble constant here

$$H = \frac{1}{2}(e_{11} + e_{22}) \quad (3)$$

and the two parameters that describe the Hubble constant anisotropy are given by

$$\Delta H_3 = e_{33} - \frac{1}{2}(e_{11} + e_{22}) \\ \Delta H_{12} = e_{11} - e_{22} \quad (4)$$

The transverse proper motions in right ascension and declination ($\mu_\alpha \cos \delta$, $\mu_\delta$) are given by

$$\mu_\alpha \cos \delta = -\frac{1}{2}(e_{11} - e_{22}) \sin 2\alpha \cos \delta = -\Delta H_{12} \sin 2\alpha \cos \delta \quad (5)$$

$$\mu_\delta = (e_{33} - \frac{1}{2}(e_{11} + e_{22})) \sin \delta \cos \delta - \frac{1}{2}(e_{11} - e_{22}) \cos 2\alpha \sin \delta \cos \delta = \\ = \frac{\Delta H_3}{2} \sin 2\delta - \frac{\Delta H_{12}}{2} \cos 2\alpha \sin \delta \cos \delta \quad (6)$$

It is obvious that if $e_{11} = e_{22} = e_{33}$ then $\Delta H_3 = \Delta H_{12} = 0$ and (2) reduces to Hubble law (1) and all proper motions described in (5)-(6) above are to be zero.

The effect of gravitational waves in the radio source proper motion has been described by Pyne et al. (1996) and Gwinn et al. (1997). It can be detected as a second degree spherical harmonic (both 'magnetic' and 'electric' type) of the expansion of the vector spherical functions (5/6 of the total effect comes to the degree 2 spherical harmonic and 1/6 - to the degree 3 spherical harmonic). If $F(\alpha, \delta)$ - a vector field of a sphere described by the components of the proper motion vector ($\mu_\alpha \cos \delta$, $\mu_\delta$)

$$\overline{F}(\alpha,\delta) = \mu_\alpha \cos\delta \cdot \overline{e}_\alpha + \mu_\delta \cdot \overline{e}_\delta \tag{7}$$

where $\overline{e}_\alpha, \overline{e}_\delta$ - unit vectors.

The same vector field $F(\alpha,\delta)$ is approximated by the vector spherical functions using the expansion

$$\overline{F}(\alpha,\delta) = \sum_{l=1}^{\infty}\sum_{m=-l}^{l} (a_{l,m}^E \overline{Y}_{l,m}^E + a_{l,m}^M \overline{Y}_{l,m}^M) \tag{8}$$

where $Y_{l,m}^E, Y_{l,m}^M$ - the 'electric' and 'magnetic' transverse vector spherical functions, correspondingly;

$$\overline{Y}_{l,m}^E = \frac{1}{\sqrt{l(l+1)}}\left(\frac{\partial V_{lm}(\alpha,\delta)}{\partial \alpha \cos\delta}\overline{e}_\alpha + \frac{\partial V_{lm}(\alpha,\delta)}{\partial \delta}\overline{e}_\delta\right)$$

$$\overline{Y}_{l,m}^M = \frac{1}{\sqrt{l(l+1)}}\left(\frac{\partial V_{lm}(\alpha,\delta)}{\partial \delta}\overline{e}_\alpha - \frac{\partial V_{lm}(\alpha,\delta)}{\partial \alpha \cos\delta}\overline{e}_\delta\right) \tag{9}$$

The function $V_{l,m}(\alpha,\delta)$ in (9) is given by

$$V_{l,m}(\alpha,\delta) = (-1)^m \sqrt{\frac{(2l+1)(l-m)!}{4\pi(l+m)!}} P_l^m(\sin\delta)\exp(im\alpha) \tag{10}$$

where $P_l^m(\sin\delta)$ - the associated Legendre functions

The coefficients of expansion (8) $a_{l,m}^E, a_{l,m}^M$ to be estimated as follows

$$a_{l,m}^E = \int_0^{2\pi}\int_{-\pi/2}^{\pi/2} \overline{F}(\alpha,\delta)\overline{Y}_{l,m}^E{}^*(\alpha,\delta)\cos\delta\, d\alpha\, d\delta$$

$$a_{l,m}^M = \int_0^{2\pi}\int_{-\pi/2}^{\pi/2} \overline{F}(\alpha,\delta)\overline{Y}_{l,m}^M{}^*(\alpha,\delta)\cos\delta\, d\alpha\, d\delta \tag{11}$$

where * means a complex conjugation.

For a finite number of points the system of equations (11) can be solved by the least squares method.

| Designation in (8) | Designation in (5) and (6) |
| --- | --- |
| $Y^E_{2,0}$ | $\Delta H_3$ |
| $Y^E_{2,1}$ | - |
| $Y^E_{2,-1}$ | - |
| $Y^E_{2,2}$ | - |
| $Y^E_{2,-2}$ | $\Delta H_{12}$ |

Table 1. Correspondence between the 'electric' spherical harmonics of second order and parameters of the Hubble constant anisotropic expansion

COMPUTATIONAL METHODS

OCCAM software (Titov, Tesmer, Boehm, 2004) analyses VLBI data by the least squares collocation method (LSCM) (Titov, 2004). The LSCM minimizes a functional similar to the conventional least-squares method and, additionally, takes into account intra-day correlations between observations. These correlations are calculated from external data. In the case of VLBI, we use the data about stochastic behavior of hydrogen clocks and wet component of troposphere delays and gradients. All estimated parameters are split into three groups on their properties: stochastic, estimated for every epoch (hydrogen clock function and wet troposphere delays), daily or 'arc' parameters to be approximately constant within a 24-hour session, and so-called 'global' parameters which are constant over the total period of observations. The reference radio source positions and the secular aberration drift were estimated as global parameters similar to the approach used by MacMillan (2003). Thirteen coefficients (3 for the first degree harmonic and 10 for the second degree harmonic) were estimated in total.

The internal motion of relativistic jets in extragalactic radio sources produces significant changes in the estimated radio source coordinates. Therefore, the positions of these astrometrically unstable quasars are treated as daily parameters to reduce the effect of instability on the estimates of other parameters. The first solution was based on the ICRF radio source classification ("defining", "candidate" and "other" sources) (Ma et al, 1998). The second solution was based on the radio source classification by Feissel-Vernier (2003) ("stable" and "unstable" sources). Though these two lists of radio sources are not independent, this approach allows assessing the sensitivity of results to variation of the list of reference radio sources. For the last two solutions all reference sources were split into the "no unstable close" (with redshift $z<1$) and " no unstable distant" ($z>1$) ones to produce independent estimates. The number of available redshifts in the existing database as large as 4.3 is limited by 352 "no unstable" radio sources therefore these last solutions are based on 172 and 150 reference radio sources with a mean redshift $z = 0.57$ and $z = 1.97$, correspondingly.

RESULTS

Estimates of the secular aberration drift components are presented in Table 2. It may be concluded that the overall average value (23 µas/year) provides a reasonable meaning for the indicated effect of secular aberration drift with the maximum 1σ standard error of 2.5 µas/year. For all four solutions the estimates of declination for all reference sources of the vector solutions are in a good agreement. However, the estimates of right ascension for two first solutions and two last solutions are different, due to a decreased number of reference radio sources in last two solutions, especially in the southern hemisphere. We consider the coordinates from two last solutions as less reliable. But it is essential that the estimates of the secular aberration drift for 'close' and 'distant' radio source are similar, i.e. this effect does not depend on distance. From two first solutions the estimated secular aberration drift vector is directed towards the point with equatorial coordinates ($\alpha$ = 265º +/- 3º, $\delta$ = 41º +/- 7º). This effect causes apparent proper motion of all quasars towards this 'focal point' with a maximum magnitude 23 µas/year (Fig 1).

The estimates of second degree spherical harmonic are less significant than for the secular aberration drift (Table 3 and 4). A deficit of radio sources in the southern hemisphere causes asymmetry of their distribution around the sky, and, eventually, large correlation between some parameters. Also these parameter estimates are sensitive to the solution strategy, for instance, when no-net-rotations (NNR) constraints are introduced to the solutions. The 3$\sigma$ standard error of 3-6 µas/year corresponds to uncertainty in the Hubble constant of 15-30 km/sec·Mpc. It exceeds the Hubble constant anomalies (10-12 km/sec·Mpc) found by McClure and Dyer (2007) from the astrophysical data.

Nevertheless, it is essential that for the 'distant' radio sources all these harmonic estimates (except the second degree zonal harmonic) are statistically significant, whereas, the same harmonics are negligible for the 'close' radio sources. It means that the unknown factor generating in the second degree electric spherical harmonics (either gravitational waves or the Hubble constant anisotropy) increases with distance to quasars.

More observations of distant radio sources in the southern hemisphere (south $\delta$ = - 40º) are required in order to make a more reliable conclusion. In addition, it is necessary to estimate the magnetic harmonics of second degree within the procedure to verify an existence of the gravitational waves.

| Solution | 1 | 2 | 3 | 4 |
|---|---|---|---|---|
| Reference radio sources | All except 102 'other' | All except 163 'unstable' | 'Close' and not 'unstable' | 'Distant' and not 'unstable' |
| Number of reference radio sources | 1559 | 1441 | 172 | 150 |
| Number of observations of reference radio sources | 2,449,601 | 2,699,600 | 1,312,924 | 1,113,180 |
| Secular aberration drift magnitude (µas/year) | 25.3 +/- 2.2 | 21.9 +/- 2.0 | 23.2 +/- 2.4 | 21.7 +/- 2.5 |
| Secular aberration drift Right Ascension | 265° +/- 3° | 264° +/- 3° | 278° +/- 5° | 287° +/- 6° |
| Secular aberration drift Declination | 41° +/- 6° | 41° +/- 7° | 43° +/- 7° | 37° +/- 9° |

Table 2. Estimates of the first degree vector harmonics with NNR-constraints for different sets of reference radio sources

| Solution | 1 | 3 | 4 |
|---|---|---|---|
| Reference radio sources (estimated as global parameters) | All except 102 'other' | 'Close' (z<1) not 'unstable' | 'Distant' (z>1) not 'unstable' |
| Number of reference radio sources | 1559 | 172 | 150 |
| Number of observations of reference radio sources | 2,449,601 | 1,312,924 | 1,113,180 |
| $a^E_{2,0}$ ($\mu as/year$) | 3.7 +/- 2.1 | 3.7 +/- 2.2 | 6.4 +/- 2.8 |
| $a^E_{2,1}$ ($\mu as/year$) | 7.4 +/- 0.9 | 1.6 +/- 1.0 | 8.2 +/- 1.5 |
| $a^E_{2,-1}$ ($\mu as/year$) | -1.9 +/- 0.9 | 0.1 +/- 1.0 | -8.2 +/-1.6 |
| $a^E_{2,2}$ ($\mu as/year$) | -0.7 +/- 0.5 | -1.4 +/- 0.6 | 4.4 +/- 0.9 |
| $a^E_{2,-2}$ ($\mu as/year$) | -2.8 +/- 0.5 | 1.6 +/- 0.6 | -1.2 +/- 0.9 |

Table 3. Estimates of the second degree vector harmonics with NNR-constraints for different sets of reference radio sources

| Solution | 1 | 3 | 4 |
|---|---|---|---|
| Reference radio sources (estimated as global parameters) | All except 102 'other' | 'Close' (z<1) not 'unstable' | 'Distant' (z>1) not 'unstable' |
| $a^E_{2,0}$ ($\mu as/year$) | 7.3 +/- 4.3 | 2.7 +/- 5.2 | 11.2 +/- 5.2 |
| $a^E_{2,1}$ ($\mu as/year$) | 7.9 +/- 1.0 | 0.6 +/- 1.3 | 5.5 +/- 1.6 |
| $a^E_{2,-1}$ ($\mu as/year$) | -1.2 +/- 0.9 | -3.2 +/- 1.3 | -6.9 +/-1.6 |
| $a^E_{2,2}$ ($\mu as/year$) | -2.2 +/- 0.6 | -1.4 +/- 1.0 | 5.0 +/- 1.1 |
| $a^E_{2,-2}$ ($\mu as/year$) | -8.0 +/- 1.3 | 1.5 +/- 1.9 | 4.1 +/- 2.1 |

Table 4. Estimates of the second degree vector harmonics without NNR-constraints for different sets of reference radio sources

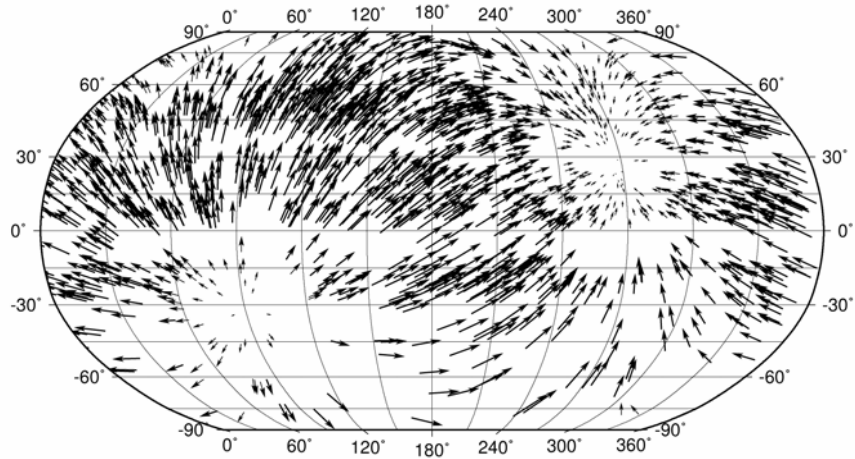

Fig 1. Vector field of the degree 1 harmonics for solution 2

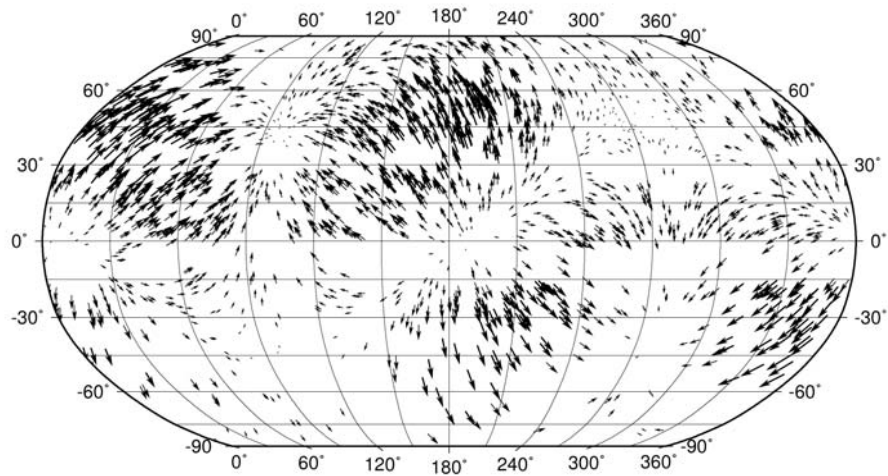

Fig 2. Vector field of the second degree harmonics for solution 1

Acknowledgments:

I am very grateful to S. Klioner for motivation of this work and useful suggestions and comments. I would like to thank D. Jauncey, G. Luton and R. Govind for careful reading of this paper and valuable comments. In addition I would like to thank M. Eubanks, D. MacMillan, M. McClure, V. Vityazev and V. Zharov for intensive discussion and useful remarks.
The paper is published with the permission of the CEO, Geoscience Australia.